\documentclass[12pt]{article}

\usepackage{graphicx}
\usepackage{amsmath}

\usepackage{booktabs}

\usepackage{multirow} 

\usepackage{tabularx}

\usepackage{caption}

\usepackage{rotating}

\usepackage{natbib}

\usepackage{url}

\usepackage{hyperref}

\usepackage{siunitx}   % For better number alignment in tables, especially S column type

\usepackage{float}

\title{Case Study: NY Real Estate Racial Equity Analysis via Applied Machine Learning}

\author{Sanjana Chalavadi, Nivea Bekal, Andrei Pastor, and Terry Leitch \\ \textit{ruxton.ai} \\ \href{mailto:terry@ruxton.ai}{terry@ruxton.ai}}

\date{May 14, 2025} % Consider using \today

\begin{document}

\maketitle

\begin{abstract}

\noindent

This study analyzes tract-level real estate ownership patterns in New York State (NYS) and New York City (NYC) to uncover racial disparities. We use an advanced race/ethnicity imputation model (LSTM+Geo with XGBoost filtering, validated at 89.2\% accuracy) to compare the predicted racial composition of property owners to the resident population from census data. We examine both a Full Model (statewide) and a Name-Only LSTM Model (NYC) to assess how incorporating geospatial context affects our predictions and disparity estimates. The results reveal significant inequities: White individuals hold a disproportionate share of properties and property value relative to their population, while Black, Hispanic, and Asian communities are underrepresented as property owners. These disparities are most pronounced in minority-majority neighborhoods, where ownership is predominantly White despite a predominantly non-White population. Corporate ownership (LLCs, trusts, etc.) exacerbates these gaps by reducing owner-occupied opportunities in urban minority communities. We provide a breakdown of ownership vs. population by race for majority-White, -Black, -Hispanic, and -Asian tracts, identify those with extreme ownership disparities, and compare patterns in urban, suburban, and rural contexts. The findings quantify persistent racial inequities in property ownership, revealing how historical and socio-economic forces are spatially manifested, and demonstrate the utility of advanced data-driven methods for diagnosing and addressing critical urban challenges.

\end{abstract}

\section{Introduction}

Homeownership is a key part of the foundation for wealth accumulation in the U.S., yet access has long been stratified along racial lines. Research in urban policy and planning has documented how historical practices like redlining and segregation created lasting racial disparities in housing equity. Today, White households have significantly higher homeownership rates and hold disproportionately more real estate wealth than minority households. In New York State (NYS), roughly 67\% of White households own their home, compared to only 33\% of Black or Hispanic households \citep{OSC2021}. These gaps are more pronounced in high-cost urban markets like New York City (NYC), where overall homeownership is only 31\% and skewed toward Whites \citep{FurmanCenterNYCHousing2023}. Such inequities have raised concerns in the urban planning literature about racial justice and housing, making it crucial to understand who owns property in communities of color \citep{Kwon2023}.

This study draws on critical urban theories to contextualize our findings. The framework of racial capitalism, which conceptualizes how racialized economic practices systematically disadvantage minority communities through housing markets and urban policy \citep{dantzler2022introduction}, is particularly salient. Gentrification frameworks, such as rent-gap theory \cite{smith1979rentgap}, help explain patterns of reinvestment and displacement often linked to shifts in property ownership. Furthermore, the financialization of housing, which transforms homes into investment assets, drives disparities in urban equity, particularly in racialized contexts \cite{aalbers2016financialization, fields2016financialization}. **Crucially, these ownership dynamics are deeply intertwined with enduring patterns of racial segregation \citep{massey1993americanapartheid} and the resultant neighborhood effects \citep{sampson2012greatamericancity}, as disparities in who owns property within a community can reinforce spatial inequalities and limit opportunities for wealth accumulation and local control for marginalized groups.**

Local data on property ownership by race is often lacking, since public records typically do not include owner race/ethnicity. To address this, we leverage machine learning for race imputation. We apply an advanced name-and-geolocation based neural network model to infer the likely race/ethnicity of property owners across NYS and NYC, allowing us to compare the racial composition of property owners to the resident population in each census tract. This builds on recent innovations in Bayesian Improved Surname Geocoding (BISG) \citep{Elliott2009} and deep learning methods for race/ethnicity prediction \citep{PastorLeitch2025}. Our approach enables a new analysis of racial ownership disparities at a neighborhood (tract) level in New York \textbf{and our machine learning-driven methodology is particularly salient for urban analysis as it provides unprecedented granularity in mapping racial ownership patterns, thereby illuminating the micro-spatial processes that underpin broader urban inequalities and informing place-based policy solutions.}

We focus on several key questions: (1) How does the racial breakdown of property owners compare to local population demographics statewide? (2) In tracts where one racial group comprises the majority (majority-White, majority-Black, etc.), what is the profile of property ownership by race? (3) Which communities show the most significant mismatches between residents and property owners, particularly tracts with high minority populations but predominantly White ownership? (4) How does the inclusion of corporate owners (e.g., LLCs, trusts, companies) influence these disparities? (5) How do patterns differ between urban, suburban, and rural areas? By answering these questions, we aim to clarify the scope and nature of racial inequities in property ownership to inform fair housing and community development policy. We highlight methodological considerations in using imputed race data, comparing a Full (name+location) model to a Name-Only model to understand imputation biases affecting disparity estimates.

Our analysis contributes to the literature on racial equity in urban planning and housing by linking data science techniques with racial justice questions. As \citet{Kwon2023} note in their review of four decades of planning research, racial equity and justice remain central to urban scholarship. This study provides new insights into racialized property ownership patterns, connecting to concerns about disinvestment, gentrification, and community autonomy. By identifying where racial ownership is widest – and how corporate investment patterns intersect – we provide planners and policymakers with information on where interventions (such as community land trusts, support for minority homebuyers, or stricter regulation of corporate landlords) are needed.

The paper is organized as follows. Section 2 describes the data sources and race imputation models (Full vs. Name-Only), and validates their accuracy and biases in a Model Ground Truth Analysis. Section 3 presents results: a statewide overview of ownership vs. population disparities, average tract profiles by majority racial group, identification of extreme-disparity tracts, impact of corporate ownership, and an urban-suburban-rural comparative analysis. Section 4 discusses the implications of these findings in light of urban policy and equity considerations, and Section 5 concludes with recommendations and future research.

\section{Data and Methods}

We employed advanced machine learning methods to impute property owners' race and ethnicity. For statewide analysis (NYS), an LSTM-based neural network integrating geolocation data and XGBoost filtering (Full Model) was utilized. For New York City (NYC), where granular geolocation data for owners was less consistently available in our dataset, a Name-Only LSTM model was applied. This dual-model approach allowed for broad coverage while acknowledging data constraints \cite{PastorLeitch2025} and for specifics on the name-only model see \citealp{Xie2021}. Detailed methodological descriptions can be found in Appendix A. For model ground truth analysis see Appendix B.

\section{Results}

\subsection{Statewide Ownership Disparities}

\begin{figure}[H]

\centering

\includegraphics[width=0.9\textwidth]{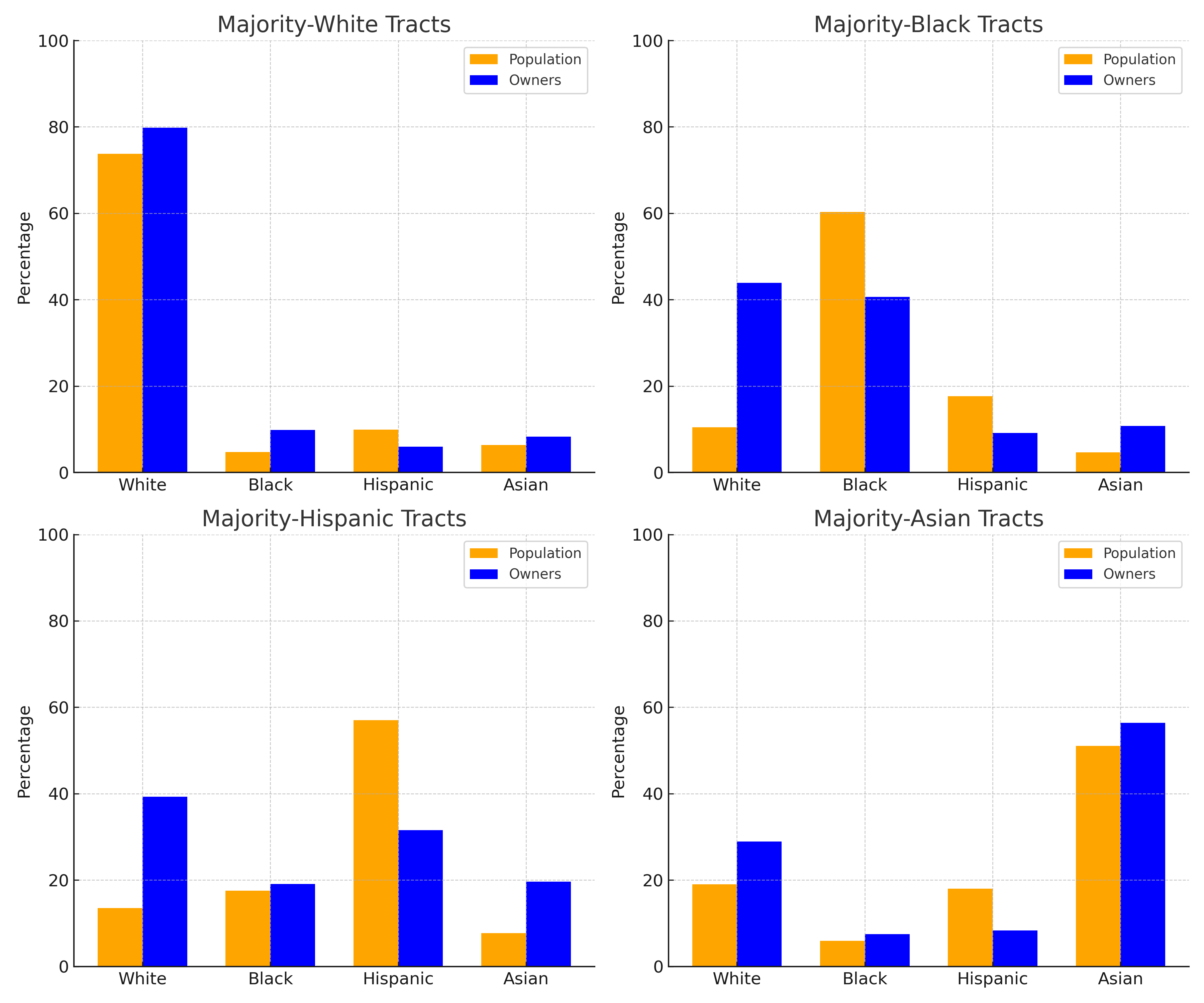}

\captionsetup{font=footnotesize}

\caption{Population vs. Predicted Owner Racial Composition in Tracts by Majority Group. Each subplot shows the average percentage of population (yellow) and of property owners (orange) that are White, Black, Hispanic, or Asian in tracts where that group is the majority. }

\label{fig:owner_vs_pop}

\end{figure}

In New York State, the racial composition of property owners often differs from the residential population, disadvantaging minority groups. Figure~\ref{fig:owner_vs_pop} illustrates this. In most tracts, the percentage of properties owned by White individuals exceeds the percentage of White residents, whereas for Black and Hispanic groups, the owner share is usually below the population share. In over 81\% of tracts, the White-ownership fraction is higher than Whites’ share of the population, indicating a significant overrepresentation of White property ownership. Conversely, Black and Hispanic residents are underrepresented among owners in many communities.

Statewide, the mean tract has 5–10 percentage points more White ownership than White population. The average tract in our dataset is about 65\% White in population but 72\% White in property ownership (unweighted). Meanwhile, the average tract is ~15\% Black by population but only ~10\% Black in ownership. Similar gaps exist for Hispanics. These imbalances reflect the pattern of higher homeownership and investment property rates for White households. Our data reveal that this pattern holds not just in aggregate, but within most local communities: even in racially mixed or minority-majority neighborhoods, property ownership tends to be whiter than the resident base, a pattern often indicative of gentrification pressures or long-standing absentee ownership, which will be explored further in the discussion.

Another way to view the disparity is through property wealth ownership. Statewide, White owners control a disproportionate share of total property market value. About 87\% of the assessed value of individually-owned real estate is held by White owners, compared to roughly 75\% of the state population. Black and Hispanic owners hold only 3–5\% of total property value, far below their population proportions (NY’s population is ~18\% Black and ~19\% Hispanic \cite{uscensus2020ny}). Asian owners hold around 8–9\% of property value, close to their ~9\% population share, though slightly under. Thus, communities of color hold significantly less property wealth than their numbers suggest. We will explore value disparities later, but even in ownership the inequity is clear.

\begin{figure}[H]

    \centering

    \includegraphics[width=\textwidth]{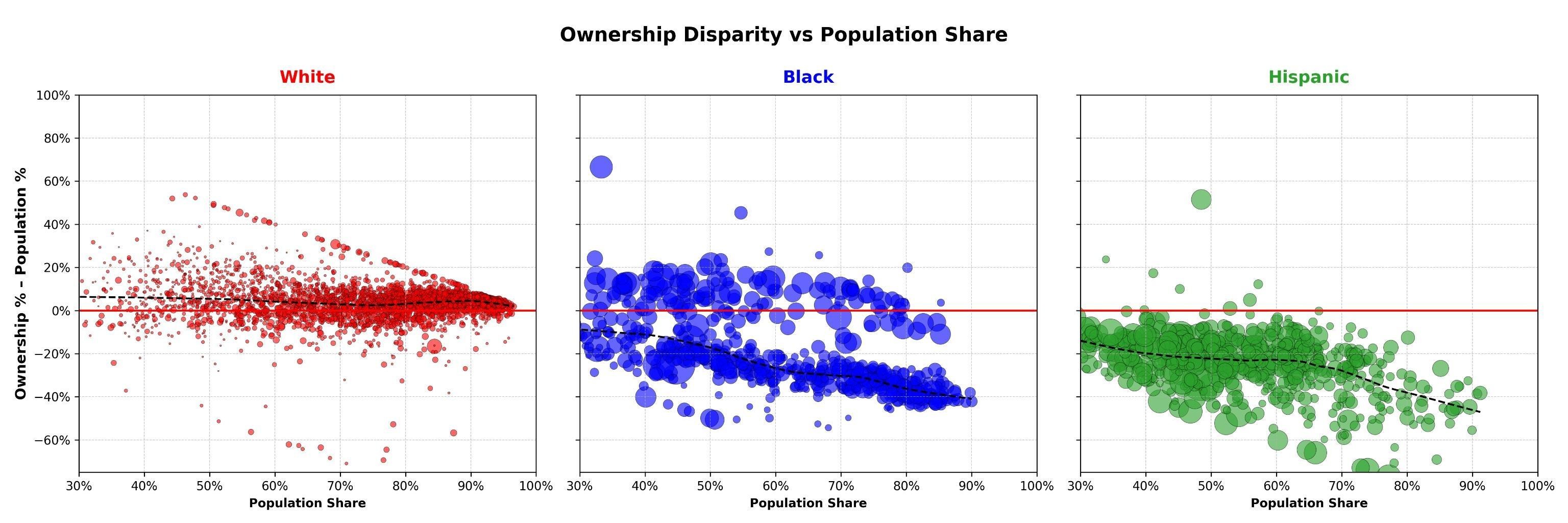}

    \captionsetup{font=footnotesize}

    \caption{Ownership Disparity (Ownership Share - Population Share) vs. Population Share by Race, Weighted by Tract Population. Each bubble represents a census tract, with size proportional to the total resident population. The dashed black line shows the LOWESS smoothed trend, and the solid red line indicates parity (zero disparity). Positive y-values indicate overrepresentation in ownership; negative values indicate underrepresentation. For Black and Hispanic populations, underrepresentation in ownership increases as their population share grows, visible across varied tracts. For the White population, overrepresentation is prevalent.}

    \label{fig:ownership_disparity_scatter}

\end{figure}

Figure~\ref{fig:ownership_disparity_scatter} illustrates these statewide patterns. For Black and Hispanic populations, the difference between their property owners and resident population is typically negative, indicating underrepresentation. This underrepresentation becomes more pronounced in tracts where these groups form a larger percentage of the population. Conversely, for the White population, property ownership share frequently exceeds their population share, particularly in tracts where they are not the majority.

Statewide trends mask neighborhood variation. We disaggregate the results by the tract’s majority racial group. Note a regional difference: disparities are more acute in the NYC metropolitan area (and other urban centers) than in rural upstate areas. In NYC, where homeownership is low, White residents are vastly overrepresented among homeowners (owning homes well above their population share) while Black and Latino residents are “severely underrepresented.” Upstate, many rural or small-town tracts are homogeneously White in population and ownership, leading to smaller gaps (due to few minorities). However, in upstate cities with sizable minority populations, we observe the same pattern of White overrepresentation. The tendency for property owners to be whiter than the population is a consistent in New York, though its magnitude varies.

\subsection{Profile of Majority-White vs Majority-Minority Tracts}

\noindent We examine average profiles for tracts categorized by racial majority. Table~\ref{tab:majority_profiles} summarizes key metrics for four tract groups: majority White, Black, Hispanic, and Asian (defined as the group $>$50\% of tract population). Approximately 3,064 tracts are majority White, 437 majority Black, 389 majority Hispanic, and 147 majority Asian (the remainder have no single group $>$50\%). The table shows the average population and owner composition by race and the average race ownership rate.

\begin{table}[h!]

\centering\footnotesize

\captionsetup{font=footnotesize}

\caption{Average Population vs. Owner Racial Composition in Majority-Race Tracts (NYS+NYC).}

\label{tab:majority_profiles}

\resizebox{.94\textwidth}{!}{

\begin{tabular}{lccccccccc}

\toprule

\multirow{2}{*}{Majority Group} & \multirow{2}{*}{Tracts} & \multicolumn{2}{c}{White (\%)} & \multicolumn{2}{c}{Black (\%)} & \multicolumn{2}{c}{Hispanic (\%)} & \multicolumn{2}{c}{Asian (\%)} \\

\cmidrule(lr){3-4} \cmidrule(lr){5-6} \cmidrule(lr){7-8} \cmidrule(lr){9-10}

& & Pop & Owners & Pop & Owners & Pop & Owners & Pop & Owners \\

\midrule

White     & 3064 & 77.8 & 81.2 & 3.8 & 7.3 & 8.6 & 4.5 & 5.2 & 6.4 \\

Black     & 437  & 6.7 & 39.3 & 69.2 & 44.6 & 14.0 & 7.2 & 3.5 & 7.8 \\

Hispanic  & 389  & 10.2 & 27.1 & 15.3 & 15.3 & 64.5 & 38.3 & 6.7 & 17.9 \\

Asian     & 147  & 15.5 & 21.6 & 2.3 & 4.0 & 15.8 & 5.7 & 63.8 & 68.1 \\

Mixed     & 888  & 28.7 & 41.4 & 19.2 & 18.6 & 26.7 & 13.8 & 18.4 & 24.0 \\

\bottomrule

\end{tabular}

}

\end{table}

Table~\ref{tab:majority_profiles} reveals clear patterns. Majority-White tracts (mostly suburban and rural) have an average population of 77.8\% White, yet their property owners are even more White (81.2\%). In these tracts, minority groups are a small share of residents and owners. Notably, Black residents are only 3.8\% of the population in the average majority-White tract, but Black owners make up 7.3\% of owners - nearly double their population share. This unexpected result (Black ownership \textit{over}representation in White areas) reflects two things: (a) the absolute numbers are very small (a difference of a few households), and (b) a few majority-White tracts have one or two Black investors who own multiple properties (e.g., a landlord from outside). Hispanic and Asian representation among owners in majority-White tracts is slightly lower than their population share (e.g. 4.5\% vs 8.6\% for Hispanics, indicating that even the small Hispanic populations in these tracts are less likely to own local property).

Majority-minority tracts show large ownership gaps. Majority-Black tracts average 69.2\% Black in population, but only 44.6\% of property owners are Black. White owners average 39.3\%, despite Whites being only 6.7\% of residents. In a typical predominantly Black neighborhood, White individuals own a disproportionate share of the properties (almost four in ten), nearly matching Black owners despite Black residents outnumbering White residents by over 10:1. This confirms a significant racial disparity: where Black families form the bulk of the community, much of the property (and control and wealth) is held by non-Black owners. Majority-Hispanic tracts show a similar pattern: population 64.5\% Hispanic, but only 38.3\% of owners are Hispanic. White owners constitute 27.1\% (versus 10.2\% White population), and Asians account for 17.9\% (versus 6.7\% population). That Asian share suggests that in many Hispanic-majority neighborhoods (e.g., parts of the Bronx or inner-ring suburbs), a significant segment of property owners may be of Asian ethnicity (reflecting investments by Asian-American landlords). Black owners in Hispanic-majority tracts are around 15\%, equal to their population share, so Blacks neither gain nor lose representation in Hispanic areas.

In NY, majority-Asian tracts, mostly in NYC (e.g., Queens and Brooklyn), show the least disparity. The population is 63.8\% Asian, and owners are 68.1\% Asian – Asians have a slight ownership advantage (+4.3 points). Whites are about 15.5\% of residents but 21.6\% of owners (moderate overrepresentation), while Hispanics are underrepresented (5.7\% owners vs 15.8\% pop). Black presence is low. These numbers suggest Asian-majority neighborhoods may have strong Asian homeownership (reflecting higher incomes or higher owner-occupancy). Whites own one-fifth of the properties through broader investment patterns, but unlike Black or Hispanic areas, the majority group (Asians) retains majority ownership. This is a distinctive dynamic of neighborhoods like Flushing or Sunset Park where ethnic community ownership is strong.

In summary, majority-White communities have higher White ownership than population (most residents and owners are White). However, majority-minority communities have much higher White ownership than White population, and the majority group (Black or Hispanic) owns substantially less. These differences translate into differences in community control and the direction of rental income and wealth: a large portion of rental payments in majority-Black neighborhoods flows to White owners elsewhere, sustaining wealth gaps.

\subsection{Extreme Disparity Tracts (`White-Owned, Minority-Populated')}

\label{sec:extreme_disparity_5tracts}

\noindent While the averages above are informative, some tracts exhibit more pronounced disparities. We define an "extreme disparity tract" as one where a racial or ethnic minority group comprises a high share of the population (i.e., White resident population is less than 50\%), yet White individuals are the predominant property owners. 

Table~\ref{tab:extreme_examples_5tracts} provides these four examples.

\begin{table}[H]

\centering

\small

\setlength{\tabcolsep}{4pt}

\caption{Four Examples of Tracts with Minority-Majority Populations and High White Ownership}

\label{tab:extreme_examples_5tracts}

\begin{tabular}{p{4.5cm} S[table-format=2.1] S[table-format=2.1] p{3cm} S[table-format=2.1]}

\toprule

\multirow{2}{=}{\textbf{Tract (Muni - Tract ID)}} & \multicolumn{2}{c}{\textbf{White}} & \multicolumn{2}{c}{\textbf{Largest Minority}} \\

\cmidrule(lr){2-3} \cmidrule(lr){4-5}

& {Pop \%} & {Indiv. Own. \%} & {Pop \% (Group)} & {Indiv. Own. \%} \\

\midrule
Brooklyn - 349.01 & 4.8 & 73.0 & Black 71.1\% & 21.3\\
Bronx - 279 & 15.7 & 41.7 & Hispanic 64.4\% & 29.8\\
Buffalo - 72.02 & 45.6 & 76.3 & Hispanic 30.3\% & 4.1\\
Syracuse - 40 & 20.0 & 40.0 & Black 37.1\% & 40.4\\
% OLD TABLE
%Brooklyn - 349.01   & 4.8 & 73.0 & Black 71.1\%  & 21.3 \\ 
%Bronx - 279         & 15.7 & 41.7 & Hispanic 64.4\% & 29.8 \\ 
%Albany - 11         & 47.9 & 41.9 & Black 32.1\%  & 45.6 \\ 
%Buffalo - 72.02     & 45.6 & 76.3 & Hispanic 30.3\% & 4.1  \\ 
%Syracuse - 40       & 37.7 & 53.5 & Black 28.4\%  & 33.5 \\ 

\bottomrule

\end{tabular}

\par\smallskip

\footnotesize{\textit{Note: ``Indiv. Own. \%'' refers to individually-owned properties.  Tract selection criteria described in text. Corresponding GEOIDs: Brooklyn - 349.01 (36047034901), Bronx - 279 (36005027900), Albany - 11 (36001001100), Buffalo - 72.02 (36029007202), Syracuse - 40 (36067004000).}}

\end{table}

These examples illustrate ownership disparities. Brooklyn tract 349.01 has 4.8\% White residents but White individuals own 73.0\% of properties. Bronx tract 279 shows 15.7\% White population with 41.7\% White individual ownership, compared to 29.8\% individual ownership by its largest minority, Hispanics (who are 64.4\% of the population). For Buffalo tract 72.02, where the White population is 45.6\%, White individual ownership is 76.3\%, significantly higher than the 4.1\% individual ownership by Hispanic residents (30.3\% of the population). Syracuse tract 40 shows White individual ownership at 40.0\% (20.0\% White Pop), compared to Black individual ownership of 40.4\% (37.1\% Black Pop).

\subsection{Stress Testing Extreme Disparity Tract Results}

\label{sec:stress_testing_extreme}

While our imputation models demonstrate good accuracy (Section 3.1), any model will have residual error. To assess the robustness of our findings regarding extreme disparity tracts, particularly concerning potential overestimation of White ownership and underestimation of minority ownership, we conduct a sensitivity analysis. We apply error rate adjustments from our PPP ground truth validation (Section 3.2, Table~\ref{tab:fpr-comparison}) to the model's predictions for the illustrative tracts.

To assess the robustness of our findings for the illustrative tracts in Table~\ref{tab:extreme_examples_5tracts}, we conduct a sensitivity analysis using error rates from our PPP ground truth validation (Table~\ref{tab:fpr-comparison} from main paper). For each tract, White individual ownership downwards using the White FPR of the model applied to its region (Full Model for Upstate: 6.9\%; Name-Only for NYC: 13.8\%). We calculate minority individual ownership upwards by dividing by (1 - FNR), using FNRs: Full Model (Black: 9.93\%, Hispanic: 15.25\%); Name-Only Model (Black: 20.62\%, Hispanic: 27.71\%).

\begin{table}[H]

\centering

\footnotesize 

\setlength{\tabcolsep}{3pt}

\caption{Four Illustrative Extreme Disparity Tracts: Model vs. Stressed Scenario Ownership (\%)}

\label{tab:stressed_disparity_tracts_final}

\resizebox{.94\textwidth}{!}{

\begin{tabular}{@{}l | S[table-format=3.2] S[table-format=3.2] | S[table-format=3.2] @{\,} p{0.2cm} S[table-format=3.2] @{\,} p{0.2cm} | S[table-format=3.2] l@{}}

\toprule

& \multicolumn{2}{c|}{\textbf{White Indiv. Own.}} & \multicolumn{4}{c|}{\textbf{Largest Minority Indiv. Own.}} & \multicolumn{2}{c}{\textbf{Resident Pop.}} \\

\cmidrule(lr){2-3} \cmidrule(lr){4-7} \cmidrule(lr){8-9}

\textbf{Tract (Muni - ID)} & {\textbf{Model}} & {\textbf{Stress}} & \multicolumn{2}{c}{\textbf{Model (Grp)}} & \multicolumn{2}{c|}{\textbf{Stress (Grp)}} & {\textbf{Maj.Min.(\%)}} & {\textbf{Group}} \\

\textbf{(Act. White Pop.)} & & & \multicolumn{2}{c}{} & \multicolumn{2}{c|}{} & & \\ 

\midrule
% TABLE 5/21/2025 AP. Correct confusion matrix
Brooklyn - 349.01 (4.8\%) & 73.0 & \textbf{62.9} & 21.3 & B & \textbf{26.8} & B & 71.1 & Black\\
Bronx - 279 (15.7\%) & 41.7 & \textbf{35.9} & 29.8 & H & \textbf{41.3} & H & 64.4 & Hispanic\\
Buffalo - 72.02 (45.6\%) & 76.3 & \textbf{71.1} & 4.1 & H & \textbf{5.7} & H & 30.3 & Hispanic\\
Syracuse - 40 (20.0\%) & 40.0 & \textbf{37.3} & 40.4 & B & \textbf{44.8} & B & 37.1 & Black\\

% TABLE 5/20/2025 AP - wrong confusion matrix
% Brooklyn - 349.01 (4.8\%) & 73.0 & \textbf{63.0} & 21.3 & B & \textbf{26.8} & B & 71.1 & Black\\
% Bronx - 279 (15.7\%) & 41.7 & \textbf{36.0} & 29.8 & H & \textbf{41.3} & H & 64.4 & Hispanic\\
% Albany - 11 (47.9\%) & 41.9 & \textbf{34.3} & 45.6 & B & \textbf{50.6} & B & 32.1 & Black\\
% Buffalo - 72.02 (45.6\%) & 76.3 & \textbf{62.6} & 4.1 & H & \textbf{4.8} & H & 30.3 & Hispanic\\
% Syracuse - 40 (20.0\%) & 40.0 & \textbf{32.8} & 40.4 & B & \textbf{44.8} & B & 37.1 & Black\\

% OLD TABLE
%Brooklyn - 349.01 (4.8\%) & 73.0 & \textbf{54.9} & 21.3 & B & \textbf{24.7} & B & 71.1 & Black \\
%Bronx - 279 (15.7\%)      & 41.7 & \textbf{31.4} & 29.8 & H & \textbf{54.8} & H & 64.4 & Hispanic \\
%Albany - 11 (47.9\%)      & 41.9 & \textbf{34.4} & 45.6 & B & \textbf{50.6} & B & 32.1 & Black \\
%Buffalo - 72.02 (45.6\%)  & 76.3 & \textbf{62.6} & 4.1 & H & \textbf{5.8} & H & 30.3 & Hispanic \\
%Syracuse - 40 (37.7\%)    & 53.5 & \textbf{43.9} & 33.5 & B & \textbf{37.2} & B & 28.4 & Black \\

\bottomrule

\end{tabular}

}

\par

\begin{minipage}{0.98\textwidth} 

\scriptsize

\textit{Note: ``Indiv. Own.'' refers to individually-owned properties. ``Model'' values from Table~\ref{tab:extreme_examples_5tracts}. ``Stress'' values calculated using White FPRs and minority FNRs detailed in text. Actual White Pop. \% from Table~\ref{tab:extreme_examples_5tracts}. Model (Grp) and Stress (Grp) refer to the largest minority for that tract from Table~\ref{tab:extreme_examples_5tracts}.}

\end{minipage}

\end{table}

Table~\ref{tab:stressed_disparity_tracts_final} shows that significant disparities persist after stress adjustments. In Brooklyn tract 349.01, stressed White ownership is 62.9\% (vs. 4.8\% White pop.), and stressed Black ownership is 26.8\% (vs. 71.1\%\ Black pop.). In Bronx - 279, stressed White ownership is 35.9\% (vs. 15.7\% White pop.), while stressed Hispanic ownership rises to 41.3\% (vs. 64.4\% Hispanic pop.), due to the high FNR for Hispanics in the Name-Only model. In upstate tracts analyzed with the Full Model, such as Buffalo - 72.02, stressed White ownership is 71.1\% versus a 45.6\% White population, and stressed Hispanic ownership is 5.7\% compared to a 30.3\% Hispanic population. This sensitivity analysis shows that while the magnitude can shift, the core finding of disparities often remains.

The finding of extreme racial disparity in property ownership persists across all illustrative tracts, even in stressed scenarios to account for misclassifications. White individuals remain the predominant property owners by a large margin, even in neighborhoods where they are a small fraction of the population. Conversely, the majority resident minority group owns a much smaller share of properties than their population share suggests, even after adjustments for undercounting. This analysis reinforces our core conclusion: significant racial inequities in property ownership are evident in these extreme disparity tracts, regardless of assumptions about imputation model error.

The authors are developing a technique to analyze the impact of errors on the conclusion drawn from the application of race classification through modeling. 

\subsection{Individual vs. Corporate Ownership}

\noindent A critical factor compounding these racial disparities is corporate ownership. Corporate or non-individual entities (LLCs, corporations, trusts, etc.) own a substantial portion of properties in New York, particularly in urban areas. Statewide, about 18.2\% of properties in our dataset are corporate-owned, with NYC higher at around 23\%, compared to about 17\% outside NYC. These entities range from small family LLCs to large real estate corporations.

Table~\ref{tab:white_corp_ownership} summarizes the effects of White individual and corporate ownership within tracts categorized by their dominant racial group. In majority-Black tracts, the average White ownership rate is 34.1\%, with corporate entities controlling an additional 21.2\%, bringing the total to 55.3\%. In majority-Hispanic tracts, this total is higher at 55.3\%, driven by high corporate ownership (39.0\%). Majority-Asian tracts show a lower total ownership rate at 38.7\%, while Mixed tracts (no dominant majority) have a total of 56.7%.

\begin{table}[H]

\centering\footnotesize

\caption{White and Corporate Ownership in Tracts by Dominant Race}

\label{tab:white_corp_ownership}

\resizebox{.94\textwidth}{!}{

\begin{tabular}{lcccc}

\toprule

Dominant Race in Tract & Tracts & White Owner (\%) & Corporate Owner (\%) & White + Corp Owner (\%) \\

\midrule
% 5/20/2025 values AP
Black & 504 & 34.1 & 21.2 & 55.3 \\
Hispanic & 500 & 16.3 & 39.0 & 55.3 \\
Asian & 155 & 21.3 & 17.4 & 38.7 \\
Mixed & 1063 & 30.5 & 26.2 & 56.7 \\

%not including white and other in this table
%White & 3200 & 64.0 & 20.1 & 84.1 \\
%Other & 6 & 20.4 & 63.3 & 83.8 \\

% OLD VALUES
%Black    & 429 & 31.3 & 21.5 & 52.8 \\
%Hispanic & 368 & 16.7 & 38.0 & 54.7 \\
%Asian    & 145 & 18.9 & 19.0 & 37.9 \\
%Mixed    & 888 & 29.9 & 26.4 & 56.3 \\

\bottomrule

\end{tabular}}

\end{table}

Corporate ownership impacts racial equity because corporations don’t have an associated race. Our racial model classifies only individual owners, thus corporate-owned properties remove housing stock from individual or household ownership. Given broader socioeconomic patterns, many corporate entities are likely White-majority owned or controlled, though we cannot attribute race directly.

Excluding corporate-owned properties and focusing on individually-owned properties, racial disparities appear less pronounced percentage-wise due to the smaller denominator. However, incorporating corporate ownership (as shown in Table~\ref{tab:white_corp_ownership}) highlights how ownership by non-individual entities reduces minority homeownership opportunities, particularly in minority-dominated tracts.

High corporate ownership occurs in majority-minority areas, especially in majority-Hispanic tracts (39.0\%) and Mixed tracts (26.2\%), compared to predominantly White areas, where it averages around 15\%. This indicates a significant portion of housing in minority neighborhoods is controlled by investors and organizations rather than resident homeowners, limiting local community control and entrenching racial economic disparities. \textbf{Such concentrations of non-individual ownership in specific urban geographies point towards targeted investment strategies that may accelerate processes of financialization, themes we revisit in the Discussion.}

Many Bronx and Brooklyn tracts exhibit high corporate ownership, diluting minority ownership with remaining individually-owned properties belonging to external White investors. In a tract with 30\% corporate ownership and predominantly White individual ownership, corporate entities significantly reduce minority ownership opportunities.

This phenomenon is pronounced in urban areas like Manhattan, where corporate ownership can exceed 60\% in some areas. This corporate presence reduces minority housing ownership to minimal levels, worsening racial disparities.

Conversely, suburban and rural areas with lower corporate ownership (often below 5\%) display disparities among individual homeowners, without significant corporate influence. Thus, corporate ownership amplifies urban racial disparities by restricting housing availability for individual and minority homeowners.

Addressing corporate dominance in minority neighborhoods could enhance minority homeownership rates. Potential interventions include programs converting corporate-held rental properties into affordable owner-occupied housing or regulating large-scale corporate property acquisitions. Without this, corporate entities will continue controlling substantial housing stock in minority communities, hindering efforts to reduce racial ownership gaps.

Corporate ownership dilutes minority ownership in diverse and minority-majority areas, worsening racial inequities in property ownership and wealth accumulation.

To illustrate this impact at a granular level, we identified tracts where the White resident population is under 50\%, but ownership by White individuals and corporate entities exceeds 50\%. Table~\ref{tab:white_corp_disparity} highlights nine such tracts, including examples from major urban areas, using our analysis data.

Table~\ref{tab:white_corp_disparity} highlights nine tracts with a White resident population under 50\%, where the combined ownership by White individuals and corporate entities exceeds 50\%. The percentages for individual racial groups are adjusted to reflect their share of all properties, determining the proportion of individually-owned (100\% - Corporate Ownership \%) and applying the model-predicted racial share of those individually-owned properties.

\begin{table}[H]

\centering\small

\caption{Nine Illustrative Tracts: Minority Population with High Combined White Individual + Corporate Property Ownership (\% of All Properties)}

\label{tab:white_corp_disparity}

\resizebox{0.99\textwidth}{!}{%

\begin{tabular}{p{4.2cm} S[table-format=2.1] S[table-format=2.1] S[table-format=2.1] p{2.8cm} S[table-format=2.1]} 

\toprule

\multirow{2}{=}{\textbf{Tract (Muni - Tract ID)}} & {\textbf{White}} & {\textbf{White Indiv.}} & {\textbf{Corp. Only}} & {\textbf{Largest Minority}} & {\textbf{Largest Minority}} \\ 

& {\textbf{Pop \%}} & {\textbf{+ Corp. Own. \%}} & {\textbf{Own. \%}} & {\textbf{Pop \% (Group)}} & {\textbf{Indiv. Own. \%}} \\

\cmidrule(lr){3-3} \cmidrule(lr){4-4} \cmidrule(lr){6-6} 

& & \multicolumn{1}{c}{\textbf{(of ALL)}} & \multicolumn{1}{c|}{\textbf{(of ALL)}} & & \multicolumn{1}{c}{\textbf{(of ALL)}} \\

\midrule
Albany - 25 (36001002500) & 15.0 & 84.1 & 79.7 & Black 58.6\% & 12.6\\
Buffalo - 72.02 (36029007202) & 45.6 & 78.7 & 9.9 & Hispanic 30.3\% & 3.7\\
Rochester - 27 (36055002700) & 2.5 & 43.5 & 33.3 & Black 78.2\% & 51.1\\
Bronx - 279 (36005027900) & 15.7 & 66.6 & 42.8 & Hispanic 64.4\% & 17.1\\
Brooklyn - 349.01 (36047034901) & 4.8 & 89.4 & 60.9 & Black 71.1\% & 8.3\\
Manhattan - 5 (36061000500) & 0.0 & 84.2 & 19.1 & Asian 40.0\% & 5.7\\
Albany - 11 (36001001100) & 47.9 & 94.1 & 89.8 & Black 32.1\% & 4.7\\
Brooklyn - 319 (36047031900) & 33.4 & 77.4 & 33.0 & Black 48.2\% & 18.3\\
Syracuse - 40 (36067004000) & 20.0 & 73.2 & 55.2 & Black 37.1\% & 18.1\\

\bottomrule

\end{tabular}

}

\end{table}

Table~\ref{tab:white_corp_disparity} illustrates the combined effect of ownership components as a percentage of total private properties in the tract. In Albany tract 11, where 47.9\% of residents are White and 32.1\% are Black, White individuals own 41.9\% of properties(Table 6), and corporations own 89.8\%. This results in a combined White individual and corporate ownership of 94.1\%. Black individuals, the largest minority group, own 4.7\% of all properties.

In Buffalo tract 72.02, 45.6\% of residents are White, and the largest minority, Hispanics, make up 30.3\%. White individuals own 76.3\% of properties, and corporations own 9.9\%, leading to a combined White and corporate control of 78.7\%. Hispanic individuals own only 3.7\% of all properties.

Table~\ref{tab:white_corp_disparity} illustrates how property control increases with corporate ownership. In Albany County tract 25, where Black residents comprise 58.6\% of the population, White individuals and corporate entities own 84.1\% of properties. Black individuals own 12.6\% of the individually-held properties. The high corporate share (79.7\%) contributes to the overall non-local/non-minority control.

Consider Erie County tract 36029007202 (Buffalo tract 72.02). Its population is 45.6\% White, 30.3\% Hispanic, and 21.4\% Black. Focusing on individual ownership, White individuals own 76.3\% of properties while the largest minority group, Hispanics, own only 3.7\%. When corporate ownership is factored in (Table~\ref{tab:white_corp_disparity}), the combined White individual and corporate ownership for this tract is 78.7\% (with corporations alone owning 9.9\%). This highlights the limited ownership stake for local minority residents in this majority non-White tract.

In Monroe County tract 36055002700 (Rochester tract 27), Black residents are 78.2\% of the population. However, combined White and corporate ownership is 43.5\%, with corporations owning 33.3\% and Black individuals owning 51.1\%. In Onondaga County tract 36067004000 (Syracuse), where Black residents are 37.1\%, they own 18.1\% of properties. Meanwhile, combined White and corporate entities control 73.2\%.

The pattern is consistent in NYC. In Kings County tract 36047034901 (Brooklyn tract 349.01), with a 71.1\% Black population (and 4.8\% White population), White and corporate ownership together account for 89.4\% of properties.

These examples show how, in many minority-majority neighborhoods, considering corporate ownership alongside White individual ownership reveals a greater concentration of property control by external or non-resident entities. This reduces the ownership stake of the local majority-minority community.

Corporate ownership illustrates how racial disparities in property control are amplified, especially in urban neighborhoods with large minority populations. In \textbf{Albany's South End} (Albany tract 11), where 32.1\% of residents are Black, White individuals and corporate entities own 94.1\% of the properties. Despite being the demographic majority, Black residents don't even own one-fifth of local housing.

In \textbf{Buffalo} (tract 72.02,), a neighborhood with a growing Hispanic population (30.3\%) and less than 46\% White residents, White and corporate ownership reaches 78.7\%, while Hispanic residents own only 3.7\% of homes. In \textbf{Syracuse} (tract 40), where Black residents account for 37.1\%, they own just 18.1\% of the properties. White and corporate entities control over 73\% of the housing stock.

The pattern is pronounced downstate. In \textbf{Brooklyn's Bedford-Stuyvesant} neighborhood (tract 349.01), where 4.8\% of the population is White and over 70\% is Black, White and corporate ownership accounts for 89.4\% of properties. In \textbf{East Harlem} (Manhattan tract 5), where Asian and Black residents dominate, the White + corporate ownership share exceeds 84\%. In the \textbf{Bronx's Soundview} area (tract 279), a 64.4\% Hispanic neighborhood, the combined White and corporate ownership figure is over 66\%.

These examples show how minority-majority neighborhoods are characterized by outsized White and institutional control of housing. Corporate ownership limits homeownership opportunities for residents of color. Because corporate entities aren’t associated with a race and are often wealthy, their presence removes housing stock from the pool of owner-occupied homes, worsening racial disparities.

This pattern is not isolated. In urban New York, high corporate ownership overlaps with minority-majority tracts, signaling a systemic issue. Corporate-controlled housing results in absentee landlords, lower reinvestment in local infrastructure, and rental income to non-local stakeholders. 

% Roadsign: Emphasize policy relevance of corporate ownership %
These findings underscore the significant role corporate entities play in limiting minority ownership and exacerbating urban racial inequalities. Corporate dominance in property markets, especially in minority-majority neighborhoods, reinforces patterns identified in financialization literature, where housing units become investment vehicles rather than homes, thereby stripping local communities of wealth-building opportunities \cite{fields2016financialization, aalbers2016financialization}. 

These findings also argue for interventions to limit speculative or absentee ownership and foster local housing equity. Approaches include pathways for tenant purchase of corporate-owned units, supporting community land trusts, or regulating large-scale corporate acquisition of housing in vulnerable neighborhoods. Without addressing this imbalance, corporate dominance will continue to undermine efforts to close the racial homeownership and wealth gaps in New York.

\subsection{Urban, Suburban, and Rural Patterns}

\noindent Finally, we examine these dynamics across the urban-suburban-rural spectrum. We use Census tract urbanization codes, classified as urban or rural based on population density and metropolitan area inclusion. About 84\% of tracts are urban, the rest rural. “Urban” includes city neighborhoods and suburban areas within metropolitan regions. We distinguish the dense urban core from suburban areas. We use property density: tracts in the top quartile of property count (generally $>$1000 properties per tract, typical of NYC and inner-city tracts) are labeled \textit{Urban-Core}, those urban by census but with fewer properties (more spread-out, often outer boroughs or suburbs) are \textit{Suburban-Urban}, and those labeled rural by census are \textit{Rural}.

Using this categorization, we observe:

\begin{itemize}

\item \textbf{Urban-Core tracts (e.g. Manhattan, inner Brooklyn):} These areas have diverse populations but heavy White ownership. Corporate ownership is highest here (often 20–30\% of properties). Manhattan census tracts with mixed race often have White owner shares 20-30 points above White population shares. Overall, urban-core tracts show the largest average racial ownership gap (mean difference White(owner) minus White(pop) $\approx +15\%$). They also have high income disparities, aligning with Argyle \& Barber’s bias scenario \citet{argyle2023}. Our Full Model mitigates some bias; the real gap remains that minorities in these areas have lower homeownership due to affordability. These dense urban cores, \textbf{characterized by their central role in regional economies and as focal points for both speculative investment and housing affordability crises,} often sites of intense capital investment and housing market pressures, exhibit the starkest racial ownership gaps.

\item \textbf{Suburban tracts (urban=Yes but lower density, e.g. Long Island, Westchester, or outer Queens/Staten Island):} In NY, these tend to be majority White, with smaller racial gaps. Many are majority-White tracts with ~85-90\% White population and owners. Some diverse suburbs (e.g., Nassau County with significant Black or Hispanic middle-class populations) show moderate gaps – e.g., a suburb with 60\% White, 20\% Black might have 75\% White, 15\% Black owners, reflecting lingering lending or income differences. On average, suburban tracts had a White owner share about 4 percentage points above White population share (compared to 15 points in urban-core). Corporate ownership in suburbs is modest (~10-15\%). Thus, they exhibit disparities but not as marked as city centers. Notably, Asian Americans have high ownership rates in some areas (e.g., Queens or Nassau where Asians are 20\% population and ~20\% of owners, sometimes slightly overrepresented, suggesting strong homeownership).

\item \textbf{Rural tracts:} These are predominantly upstate areas, often $>$90\% White. In these tracts, ownership is almost entirely White (95-100\%). There is little racial gap due to few minorities. In a few rural tracts with notable minority populations (e.g., a Native American reservation or a prison with a diverse inmate population), we saw cases of near-zero local minority property ownership (since minorities might be transient or low-income). Generally, rural NY has high homeownership and low diversity, so the “percent gap” measure is small. They have a high absolute gap in that White families own almost all the land and non-Whites almost none, but given the demography, this is expected. Corporate ownership is lowest in rural tracts (~5-8\% on average), mostly farms or large land parcels owned by companies.

\end{itemize}

Comparing NYC (urban) to upstate: In NYC, only 31\% of households own homes (ACS data) and those owners are mostly White. Our tract analysis reflects this: in many NYC tracts, especially with majority Black or Hispanic populations, the few homeowners are largely White. Upstate (including suburbs), homeownership rates are higher (~60\% on average), and many owners are local (due to affordable housing). Yet, upstate cities show a pattern like NYC’s: in Buffalo’s East Side (majority Black), many properties are owned by White suburban landlords. In small towns, if integrated, White residents have higher ownership rates than Black residents (due to income gap or recent move-ins).

Integrated suburban tracts (no majority group) often have interesting splits. We noted NYC tracts with balanced populations (e.g., 40\% White, 30\% Black, 20\% Hispanic, 10\% Asian) where ownership was skewed (e.g., 60\% White owners, 20\% Black, 15\% Asian, 5\% Hispanic). Those in Queens or Westchester highlight that in a mixed community, minority groups lag in homeownership relative to Whites and Asians. This is due to differences in median income or immigration status (Asian households in NY have higher median income than Black or Hispanic, leading to higher homeownership).

To summarize urban vs rural: In dense urban neighborhoods (often minority-majority), corporate ownership compounds racial ownership disparities. Suburban areas have moderate disparities, and rural areas the least (but least diverse). These patterns align with observations that urban housing markets, with higher prices and more investors, create barriers for local minority homeownership, whereas rural areas, though more accessible, have smaller minority populations. Suburban zones fall in between; they reflect post-war housing policies that enabled significant White homeownership (and excluded others via redlining), effects that persist in current homeownership.

\section{Discussion}

The results reveal significant racial disparities in property ownership. Despite progress since mid-20th century housing discrimination, the legacy remains. Predominantly minority neighborhoods in the Bronx or Buffalo exhibit low homeownership rates and substantial outside (often White) ownership. Consequently, when neighborhood properties appreciate, economic benefits flow to these external owners, rather than to minority residents who become "renters in their own communities." Our findings align with prior research highlighting capital extraction from Black and Brown communities through absentee ownership and speculative investments \citealp{Taylor2019RaceForProfit,fields2021}.

Racially uneven homeownership has excluded many minority households from a key wealth accumulation avenue: real estate capital gains. Since major government interventions promoting homeownership began—including federally backed mortgages via the Federal Housing Administration (FHA) and GI Bill benefits favoring White households—cumulative real estate capital gains have reached an estimated \$30–40 trillion nationwide \citep{NAR2024HomePrice,ZillowHousingValue2023,Census1940HomeValue}. Approximately \$2–3 trillion generated in New York State alone \citep{NYSAR2021MedianPrice,FurmanCenterNYCHousing2023}. The racial disparities in property ownership create significant inequalities in wealth accumulation opportunities for minority communities.

These gains have mostly benefited households to purchase homes early and hold them long-term. These groups have historically skewed White due to discrimination and barriers in lending, zoning, and credit access. Exclusion from this property appreciation cycle restricts opportunities for Black, Hispanic, and other marginalized groups to build intergenerational wealth, widening racial wealth gaps and economic inequality in the U.S.

In urban planning and policy, these disparities raise questions about community control, gentrification, and displacement. Extreme cases in Brooklyn or Harlem are neighborhoods undergoing rapid change. White ownership of formerly minority-owned housing signals gentrification, though in some cases, it has been the norm for decades, as with slumlords in segregated areas. Understanding the extent of external ownership could help target anti-displacement measures. The City of New York might use this analysis to identify neighborhoods where tenant protections or incentives for first-time homebuyers should be strengthened to prevent further loss of community ownership.

A key discussion point is corporate ownership. Our data confirm the significant challenge posed by “corporate landlords” in \textbf{New York City}, a phenomenon whose proliferation and impact are detailed in comprehensive analyses like \citep{LISC2022GamblingWithHomes}, often reflecting issues raised by community advocates. This trend, also documented nationally (e.g., rise of institutional single-family rental investors \citep{goodman2023profile}), can exacerbate affordability issues and reduce the likelihood of home purchase by local families. Policymakers could consider strategies to limit speculative corporate purchases in affected tracts – e.g., taxes on LLC-owned residential properties or programs to transfer housing to community land trusts or owner-occupants. One proposed solution is giving tenants the right to purchase (Tenant Opportunity to Purchase or TOPA) when a landlord sells, as advocated by \citep{tegeler2020topa,haberle2020call}. This could convert rental housing into owned homes for existing residents, increasing minority ownership.

Another point of discussion is the difference between the Full Model and Name-Only Model results. While the broad conclusions remain consistent, the potential underestimation of minority owners in NYC due to name-only predictions suggests caution. We interpreted patterns qualitatively and in aggregate. Future work will incorporate additional data (such as voter registration or ethnicity surveys) to validate the predictions in NYC. This highlights a broader methodological point: using algorithms to infer race can introduce biases if not carefully designed. In our study, we were concerned with not overestimating minority ownership (a false negative for disparity). The Full Model’s lower White FPR helped ensure we didn’t mistakenly count too many owners as White. The Name-Only model likely slightly overestimates White ownership in some tracts, meaning the real disparities could be larger than reported for NYC. Thus, our conclusions are cautious.

Our findings align with existing literature. A national analysis by \citet{howell2022} found that Black neighborhoods have lower rates of institutional investment in homeownership and that homes in these neighborhoods are devalued relative to similar homes in White neighborhoods (the appraisal gap). While our focus is different (ownership vs. population rather than valuation), the concept that minority communities face structural disadvantages in property ownership is reinforced. \citet{Kwon2023} emphasize how planning research has focused on equity and justice; our data-driven approach provides concrete metrics. It quantifies inequity not just in ownership rates but at the neighborhood level where planning interventions occur.

One limitation is that we did not differentiate between owner-occupied and investor-owned properties. A White individual owning a home in a Black neighborhood might be an owner-occupant (e.g., a gentrifier) or an absentee landlord. Our data treat both as “White owner.” If we had data on the owner’s address, we could separate local vs absentee owners. That would add another dimension: how much of the disparity is due to gentrification (different race homeowners) versus absentee landlordism. Anecdotally, both phenomena are present. In up-and-coming areas (say, Harlem), there’s an influx of non-Black homeowners (thus White or Asian owner-occupants). In disinvested areas (say, parts of the Bronx), there are mostly renters and externally located owners. Differentiating these would require further analysis.

We noted median income differences among tract categories. Majority-White tracts had much higher median incomes than majority-Hispanic tracts, 90k vs 62k. This contributes to the disparities: lower-income populations have less access to mortgages and capital. We correlated income with ownership gaps: tracts with lower median income have larger ownership gaps (low-income usually means minority or mixed with low ownership). Previous studies showed socioeconomic status correlates with BISG misclassification \citep{argyle2023}, capturing the same pattern: affluent areas have more owners who are often White. Our approach allows those structural socioeconomic effects to be observed as outcomes (disparities), which are significant for equity.

Our results also have direct implications for housing policy and urban planning in New York State. Recent reports by the Department of Financial Services (DFS) \cite{dfs2024minoritylending, dfs2022redlining} and Governor Hochul’s housing initiatives emphasize the need for reforms targeting lending practices and corporate property acquisitions in minority-majority neighborhoods. Policies such as enhanced monitoring of corporate purchases, regulatory adjustments to the Community Reinvestment Act (CRA), and incentives supporting minority homeownership could mitigate the disparities documented in our analysis. For example, mandating disclosures of beneficial corporate ownership, combined with proactive monitoring using advanced analytical methods similar to those employed here, can form part of a broader policy strategy to foster racial equity in urban housing markets.

One key limitation, acknowledged earlier, is our current inability to systematically differentiate between White owner-occupiers and White absentee landlords/investors within minority-majority tracts. This distinction is critical for urban theory and policy. For instance, an influx of White owner-occupiers into a historically minority neighborhood might be an indicator of gentrification, potentially leading to rising property values that benefit new entrants but also create displacement pressures on long-term, often lower-income, residents of color \citep{smith2002newglobalism, lees2008gentrification}. Such dynamics alter the social and economic fabric of neighborhoods in ways distinct from those shaped by absentee landlordism.

Conversely, high levels of White absentee ownership, particularly if long-standing, may signify patterns of disinvestment in the local community, extraction of rental income without commensurate reinvestment in property upkeep or local amenities, and a weakening of local civic engagement \citep{Taylor2019RaceForProfit}. While both scenarios contribute to the racial wealth gap and limit community control for residents of color, the urban processes and appropriate policy responses differ. Future research, ideally leveraging owner address data or more sophisticated proxies for owner-occupancy, is essential to disentangle these varied forms of external ownership and their specific impacts on neighborhood change, racial equity, and the lived experiences of urban residents.

\section{Conclusion}

This study examined racial disparities in property ownership across New York State using predictive analytics. By comparing the predicted racial makeup of property owners to local demographics, we identified mismatches overlooked by traditional data sources. The results confirmed that, statewide, White individuals are disproportionately represented among property owners, holding far more real estate (number and value) than their population share. In contrast, Black and Hispanic New Yorkers are underrepresented among property owners, even in neighborhoods. Asian ownership rates are closer to parity, though context-dependent.

In majority-Black and majority-Hispanic communities, less than half of properties are owned by residents of the same racial/ethnic group. Many are owned by White individuals, indicating economic control at odds with the community’s racial identity. These ownership gaps reflect and reinforce deeper inequities in wealth and political power. Lower homeownership means less community stability and fewer opportunities for minority families to build wealth across generations.

Our analysis highlighted corporate ownership in urban areas as a barrier to increasing homeownership in minority communities. Policies to improve racial equity in housing need to address individual discrimination and the broader market forces of investment and speculation targeting low-income, minority neighborhoods. A worthwhile follow-up study would be a longitudinal analysis of corporate ownership via multiyear tax data.

We demonstrated the utility of an AI-enhanced public policy analysis approach that combines data from disparate sources and employs machine learning to infer demographic characteristics. It can be applied to other cities and states to uncover hidden patterns of ownership or lending discrimination. We underscored the importance of model choice. A Full Model incorporating geographic context provided more reliable results by reducing racial misclassification bias, whereas a simpler Name-Only Model, while useful, requires caution. Future work could integrate additional validation using self-reported race data for a subset of owners to refine predictions.

A weakness in applying models is that the predicted population may differ from the model’s origin. To compensate, we used an out-of-sample dataset overlapping with the population. We used error rates from that analysis to show that extreme disparity cases using conservative assumptions still held. The authors conceived a technique for estimating the error distribution for prediction models to share in a forthcoming paper that better encapsulates the error impact across all scenarios, not just extreme disparities. 

Policy attention is needed to address racial disparities in property ownership. Strategies to increase homeownership among Black and Hispanic households – such as down-payment assistance, subsidized mortgages, or community land trusts – could help close the gap. Efforts to curb speculative ownership and support tenant-to-owner pathways in impacted neighborhoods may also be critical. This research quantifies the problem, providing a foundation for measuring progress. As plans and interventions are implemented, the same data-driven lens can track changes in ownership patterns yearly. Achieving racial equity in housing will require reversing the documented patterns, ensuring home or investment property is accessible to all racial groups, not just the historically privileged.

Future research should extend our methodological framework longitudinally and deepen the theoretical exploration of how racial capitalism and financialization manifest spatially through corporate ownership, gentrification, and displacement. Incorporating theoretical insights from racial capitalism and financialization, subsequent studies can more robustly explore how housing policies and market dynamics interact, thereby guiding effective urban planning interventions that prioritize racial equity.

\bibliography{references}

\appendix
\section{Appendix A: Technical Details of Race/Ethnicity Imputation Model}
\subsection{Property and Demographic Data}

Our analysis integrates parcel-level property data with census tract-level demographic data. We obtained the property records from NYS \citealp{NYSASSESSDATA} and NYC \citealp{NYCTAXDATA}  (property tax assessments and deed registries). The records contain each property’s location, assessed value, and the owner’s name and ownership type (individual or entity). The dataset includes millions of parcels across all 62 counties of NYS, including NYC’s five boroughs. We geocoded each property to its 11-digit census tract (2020 boundaries) and attached tract-level population demographics from the 2020 Census and American Community Survey (ACS). Key census variables include total population and racial/ethnic composition (Non-Hispanic White, Non-Hispanic Black or African American, Hispanic/Latino, Non-Hispanic Asian, and others). We also incorporated tract median household income from the ACS for socio-economic context \citealp{CensusGeocoder}.

We classified each property record by owner type using a Natural Language Processing (NLP) script to parse owner names and assign an owner type code to each property record. The owner type codes are: 1 = individual (person or family name) vs. 2,3,4 = non-individual (corporations, LLCs, trusts, government, etc.). Names with suffixes like ``LLC'', ``Inc.'', or ``Trust'' were labeled as corporate entities. This allows us to distinguish individual homeowners from institutional investors or corporate landlords. We exclude any tract with fewer than 100 total property records to ensure reliable percentage estimates.

After cleaning and filtering, we aggregated the data by census tract. For each tract, we computed the number of properties and sum of property value owned by each owner racial group (predicted by our model) and owner type. We calculated the fraction of properties (and property value) owned by corporate entities in each tract. These aggregates form the basis of our comparison between the racial composition of property owners and the tract’s resident population.

\subsection{Race/Ethnicity Imputation Models}

We apply two predictive models to infer property owners’ race/ethnicity: (1) a \textbf{Full Model} that integrates name and geolocation features, and (2) a \textbf{Name-Only Model} using only the owner’s name. Both were developed in prior work and are variants of an LSTM (Long Short-Term Memory) neural network trained on labeled voter registration data (from Florida and North Carolina) for race/ethnicity \citealp{PastorLeitch2025}.

The Full Model, an LSTM+Geo+XGBoost ensemble detailed in \citet{PastorLeitch2025}, extends a character-level name LSTM by incorporating census tract demographics as an additional input (like Bayesian Improved Surname Geocoding but learned jointly via the LSTM). It also includes a post-processing step: the LSTM+Geo outputs are fed into an XGBoost decision tree to refine predictions and reduce bias. The Name-Only Model is the baseline character-level LSTM using only the name string (first,last name tokens). We applied the Full Model to statewide data (NYS tracts including suburbs and upstate cities) and the Name-Only Model to NYC data. This choice was motivated by data availability and the desire to assess model differences: the Full Model required geolocation priors for upstate owners, whereas NYC was used as a test-bed for the name-only approach.

Both models output a probability distribution over five racial/ethnic categories for each owner: White (Non-Hispanic), Black/African American (Non-Hispanic), Hispanic/Latino, Asian, or Other (including Native American, multiracial, etc.). We assign each owner to the race with the highest predicted probability.

\subsection{Validation and Model Bias Analysis}

Before presenting results, we assess the reliability of the imputed race data by examining model performance and potential biases. We focus on the differences between the Full Model (NYS) and Name-Only Model (NYC). The Full Model’s geolocation incorporation mitigated the bias identified by \citet{argyle2023}, who found that standard BISG misclassifies high-SES minorities as White and low-SES Whites as minority. Our Full Model’s bias-corrective approach (LSTM+Geo with XGBoost filtering) achieves a lower White misclassification rate. In prior validation, the LSTM+Geo+XGBoost Full Model had about 89.2\% overall accuracy and reduced bias compared to traditional BISG \citealp{argyle2023}. The model’s false positive rate (FPR) for labeling someone White (misclassifying a non-White person as White) was only 17.8\%, much lower than BISG’s 41.8\% or a name-only LSTM’s 24.6\% \citep{PastorLeitch2025}. This bias reduction is significant, as misidentifying minority owners as White would obscure the disparities we aim to measure. The Name-Only Model, lacking geographic context, misclassifies minorities in affluent or majority-White areas as White. Thus, our NYC results (from the name-only classifier) may underestimate minority ownership in mixed or gentrifying tracts. We address this limitation by comparing patterns between the two model outputs.

\appendix
\section{Appendix B: Model Ground Truth Analysis}

\label{sec:model-ground-truth-analysis}

We conducted a ground truth evaluation using the Paycheck Protection Program (PPP) dataset \citep{SBA_EIDL_Data,SBA_PPP_Data} for New York State to assess two model configurations: the \textbf{Full Model} (LSTM with geolocation and XGBoost filtering) and the \textbf{Name Only Model} (LSTM without geolocation or XGBoost filtering). Each PPP loan applicant provided self-identified race and ethnicity data, offering a strong benchmark. After preprocessing, the analysis dataset contained \num{35134} validated loans.

\subsection{Overall Model Performance}

Table~\ref{tab:model-comparison-metrics} summarizes performance metrics for each model, evaluated against PPP ground truth.

\begin{table}[H]

  \centering

  \begin{tabular}{lS[table-format=1.4]S[table-format=1.4]S[table-format=1.4]S[table-format=1.4]}

    \toprule

    \textbf{Model} & {\textbf{Accuracy}} & {\textbf{WtAvg Precision}} & {\textbf{WtAvg Recall}} & {\textbf{WtAvg F1}}\\

    \midrule

    Full Model & 0.8725 & 0.8701 & 0.8725 & 0.8705\\

    Name Only Model & 0.8174 & 0.8203 & 0.8174 & 0.8167\\

    \bottomrule

  \end{tabular}

  \caption{Comparison of performance metrics between Full and Name Only models using NY PPP ground truth.}

  \label{tab:model-comparison-metrics}

\end{table}

The Full Model significantly outperformed the Name Only Model across all metrics, emphasizing the importance of geolocation data and ensemble learning (XGBoost) for accurate predictions.

\subsection{Error Structure Analysis}

Confusion matrices and per-class false-positive rates (Tables~\ref{tab:full-model-confusion}, \ref{tab:name-only-confusion}, and \ref{tab:fpr-comparison}) reveal patterns of prediction errors.

\begin{table}[H]

  \centering

  \scriptsize 

\captionsetup{font=footnotesize}

  \caption{Confusion matrix for the Full Model. Rows indicate True Class; Columns indicate Predicted Class.}

  \label{tab:full-model-confusion}

\resizebox{.9\textwidth}{!}{%

  \begin{tabular}{@{}l*{5}{S[table-format=5.0]}@{}} % S column for numbers

    \toprule

    & \multicolumn{5}{c}{\textbf{Predicted Class}} \\

    \cmidrule(lr){2-6}

    \textbf{True Class} & {white5} & {afrAmer5} & {hisp5} & {asian5} & {other5}\\

    \midrule

    white5   & 10282 & 669 & 278 & 172 & 29\\

    afrAmer5 & 788   & 11344 & 270 & 141 & 44\\

    hisp5    & 438   & 625 & 2762 & 77 & 3\\

    asian5   & 321   & 270 & 40 & 6255 & 69\\

    other5   & 86    & 118 & 16 & 25 & 12\\

    \bottomrule

  \end{tabular}

}

\end{table}

\begin{table}[H]

  \centering

\normalsize

  \scriptsize % Kept as original

\captionsetup{font=footnotesize}

  \caption{Confusion matrix for the Name Only Model. Rows indicate True Class; Columns indicate Predicted Class.}

  \label{tab:name-only-confusion}

\resizebox{.9\textwidth}{!}{%

  \begin{tabular}{@{}l*{5}{S[table-format=5.0]}@{}} % S column for numbers

    \toprule

    & \multicolumn{5}{c}{\textbf{Predicted Class}} \\

    \cmidrule(lr){2-6}

    \textbf{True Class} & {white5} & {afrAmer5} & {hisp5} & {asian5} & {other5}\\

    \midrule

    white5   & 9920 & 1002 & 356 & 120 & 32\\

    afrAmer5 & 2051 & 9991 & 334 & 185 & 26\\

    hisp5    & 504  & 510 & 2823 & 67 & 1\\

    asian5   & 621  & 217 & 63 & 5982 & 72\\

    other5   & 98  & 114 & 16 & 25 & 4\\

    \bottomrule

  \end{tabular}

}

\end{table}

We provide the calculated FPR rates for both models across all races from the confusion matrices. These are used later to generate a stressed scenario to check the robustness of our results to prediction errors.

\begin{table}[H]

  \centering

  \captionsetup{font=footnotesize}

  \caption{Comparison of false-positive rates (FPR) by racial class for each model.}

\resizebox{.9\textwidth}{!}{%

  \begin{tabular}{lS[table-format=1.4]S[table-format=1.4]}

    \toprule

    \textbf{Class} & {\textbf{Full Model FPR}} & {\textbf{Name Only Model FPR}}\\

    \midrule

    white5   & 0.069 & 0.138\\

    afrAmer5 & 0.075 & 0.082\\

    hisp5    & 0.019 & 0.025\\

    asian5   & 0.015 & 0.014\\

    other5   & 0.004 & 0.004\\

    \bottomrule

  \end{tabular}

  }

  \label{tab:fpr-comparison}

\end{table}

The Full Model consistently showed lower false-positive rates across almost all classes, highlighting its ability to reduce misclassification bias.

\subsection{Socio-economic Disparities}

\paragraph{Model FPR by Income Decile}

\begin{figure}[H]

  \centering

  \includegraphics[width=0.7\textwidth]{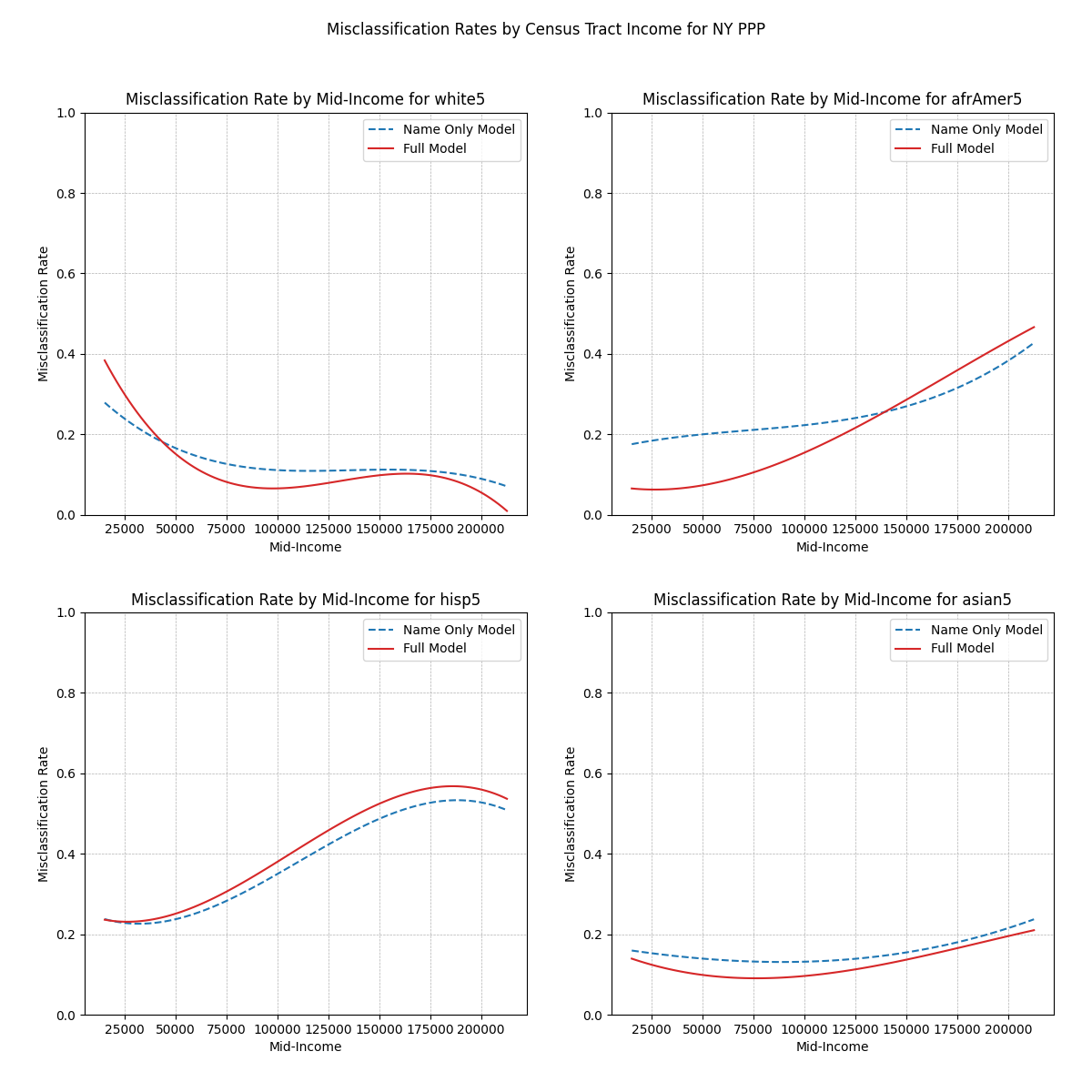}

\captionsetup{font=footnotesize}

  \caption{Accuracy by census-tract income decile for the Full Model and Name Only Model.}

  \label{fig:accuracy-income-comparison} % Changed label for clarity if this is a combined plot

\end{figure}

\noindent The Name Only Model showed slight accuracy degradation across incomes compared to the Full model overall except for where African Americans earn less than \$100k income where it underperformed the full model by 10-12\% Figure~\ref{fig:accuracy-income-comparison} 

\paragraph{Implications for Real Estate Study}

These results clarify accuracy and bias limitations, as NYC real estate data lacks explicit addresses, restricting analysis to the Name Only Model for the city. While it maintains acceptable predictive performance overall, particularly within mid-income ranges, users must interpret results cautiously due to increased misclassification errors, especially in NYC boroughs. Conversely, the availability of address data outside NYC allows the Full Model’s superior accuracy and fairness for broader NY state analyses.

\end{document}